\documentclass[aps,twocolumn]{revtex4-2}
\usepackage[bookmarksnumbered,bookmarksopen,colorlinks,citecolor=red,linkcolor=blue]{hyperref}
\usepackage{graphicx}
\usepackage{bm}
\usepackage{amsmath}
\usepackage{color}
\usepackage{booktabs}
\usepackage{multirow}
\usepackage{diagbox}

\graphicspath{{./figs/}}

\begin{document}
\title{Fully-heavy baryons $QQQ$ in vacuum and hot QCD medium }
\author{Jiaxing Zhao$^a$}
\email{jzhao@subatech.in2p3.fr}
\author{Shuzhe Shi$^b$}
\email{shuzhe-shi@tsinghua.edu.cn}
\affiliation{$^a$SUBATECH, Universit\'e de Nantes, IMT Atlantique, IN2P3/CNRS, 4 rue Alfred Kastler, 44307 Nantes cedex 3, France\\
      	 $^b$Physics Department, Tsinghua University, Beijing 100084, China}
\date{\today}
\begin{abstract}
We study the properties of fully-heavy baryons in the vacuum and the hot QCD medium, which is created in relativistic heavy-ion collisions. Masses and wave functions of $\Omega_{ccc}$, $\Omega_{ccb}$, $\Omega_{bbc}$, and $\Omega_{bbb}$ up to the second radial excited states are obtained by solving the three-body Schr\"odinger equation with Hyperspherical Harmonics method. 
With parameters completely fixed by fitting quarkonium boundstates in vacuum, we predicted the masses for $1S$, $2S$, and $3S$ states of fully-heavy baryons.
We also computed the temperature dependence of baryon masses and the thermal widths in a hot QCD medium. These properties are important to precise study of fully-heavy baryon production in heavy ion collisions.
\end{abstract}

\maketitle

\section{introduction}
\label{section1}

Taking into account the fact that charm and bottom quarks are very heavy and their moving velocity is small, there exists a hierarchy of scales in the study of heavy quarks: $m\gg mv \gg mv^2$~\cite{Caswell:1985ui,Brambilla:1999xf}. Integrating out the degrees of freedom with momenta larger than $m$ and $mv$ successively in the QCD Lagrangian, one can derive its nonrelativistic versions NRQCD and pNRQCD~\cite{Brambilla:1999xf}. Furthermore, if neglecting the color transition between the color-singlet and color-octet states, the pNRQCD becomes a potential model~\cite{Brambilla:1999xf}. In this case, one can employ the Schr\"odinger equation to study the properties of hadrons consisting of only heavy quarks. It turns out that the mass spectra of quarkonium are explained very well based on the Schr\"odinger equation with Cornell potential, see the review paper~\cite{Zhao:2020jqu}. The Schr\"odinger equation has also been extended to three-body case to predict the masses of fully-heavy baryons~\cite{Silvestre-Brac:1996myf,Martynenko:2007je,Vijande:2015faa,Liu:2019vtx,Yang:2019lsg,Shi:2019tji}.

Searching for the QCD phase transition is one of the physics goals of relativistic heavy-ion collisions. The phase transition happens around $160$~MeV predicted by the lattice QCD and effective models~\cite{Bazavov:2011nk, Fodor:2004nz, Xin:2014ela, Li:2018ygx}. The high temperature phase is the so called quark-gluon plasma (QGP) and has been confirmed in experiments~\cite{Akiba:2015jwa}. This hot QCD medium changes quarkonium production rate, compared to the vacuum case which happens in $pp$ or $e^+e^-$ collisions.
Quarkonium suppression was proposed as a smoking gun of the production of the QGP~\cite{Matsui:1986dk}. The evolution and production of quarkonium in QGP medium depends closely on their finite-temperature properties, such as binding energy and width~\cite{Blaizot:2015hya, Yao:2020eqy, Brambilla:2020qwo, Miura:2022arv, Villar:2022sbv, Lafferty:2019jpr}. Analogously, to study the yield of fully-heavy baryons $QQQ$ in the QGP, their finite-temperature properties are demanded. 
Many theoretical studies show the $\Xi_{cc}$ and $\Omega_{ccc}$ yields per binary nucleon-nucleon collision in heavy-ion collisions at RHIC and LHC will be largely enhanced in comparison with nucleon-nucleon collisions in vacuum~\cite{He:2014tga, Zhao:2016ccp, Cho:2019syk, Andronic:2021erx}. Searching for these fully-heavy baryons in relativistic heavy ion collisions at LHC energy attracted many attentions and has been listed as one of goal of next-generation LHC heavy-ion experiment~\cite{Adamova:2019vkf}. Therefore, it is important to figure out their properties in the hot QCD medium, especially with the help of the newly obtained heavy quark finite-temperature potential~\cite{Burnier:2015tda, Bala:2021fkm}. 
In this work, we employ the three-body Schr\"odinger equation to study the properties of fully-heavy baryon states $\Omega_{ccc}$, $\Omega_{ccb}$, $\Omega_{bbc}$, and $\Omega_{bbb}$ in both the vacuum and finite temperatures. 

The structure of this paper is as follows. In Sec.~\ref{section2} we present the framework of solving the three-body Schr\"odinger equation. The baryon properties, including mass and size, in the vacuum and a hot medium, are investigated in Secs.~\ref{section3} and~\ref{section4}, respectively. A summary is given in Sec.~\ref{section5}.

\section{Theoretic framework}\label{sec.framework}
\label{section2}
\subsection{Coordinate Transformation}
For a system of three quarks with the mass $\mu_i(i=1,2,3)$, the wave function $\Psi({\bf r}_1,{\bf r}_2,{\bf r}_3)$ and the energy $E$ satisfy by the Schr\"odinger equation
\begin{equation}
\left( \sum_{i=1}^3 {\hat {\bf p}^2_i\over 2\mu_i}+ \sum_{i<j}V_{ij}(|{\bf r}_{ij}|)\right)\,\Psi = E\,\Psi,
\end{equation}
under the boundary condition that the wave function vanishs when coordinates approaches to infinity. We have neglected the direct three-body potentials and assumed that the interaction potential is the summation of the two-body interactions. Taking into account one-gluon-exchange interaction, the two-body potential can be effectively expressed as~\cite{Wong:2001td,Kawanai:2011jt},
\begin{equation}
V_{ij}(|{\bf r}_{ij}|) = -{\lambda_i^a\cdot\lambda_j^a\over 4}\left(V^c_{ij}(|{\bf r}_{ij}|)+V^{ss}_{ij}(|{\bf r}_{ij}|){\bf s}_i\cdot{\bf s}_j\right),  
\label{v2}
\end{equation}
where $\lambda_i^a\ (a=1,...,8)$ are the SU(3) Gell-Mann matrices, the factor $1/4$ is from the normalization, $V^c_{ij}$ is the spin independent interaction, $V^{ss}_{ij}$ is the strength of the spin-spin interaction, and $|{\bf r}_{ij}|=|{\bf r}_i-{\bf r}_j|$ is the distance between two quarks labeled by $i$ and $j$. 
We employ the Cornell potential to describe the spin-independent central interaction $V_{ij}^c$ between two quarks and the lattice result~\cite{Kawanai:2011jt} for the spin-spin coupling, 
\begin{eqnarray}
V^c_{ij}(|{\bf r}_{ij}|) &=& -{\alpha \over |{\bf r}_{ij}|}+\sigma |{\bf r}_{ij}|, \nonumber\\
V^{ss}_{ij}(|{\bf r}_{ij}|) &=& \beta e^{-\gamma |{\bf r}_{ij}| }.
\label{eq.cornell}
\end{eqnarray}
The parameters in the potential like $\alpha$, $\sigma$, $\beta$, and $\gamma$ are given by lattice QCD with some uncertainties, which can be fixed by fitting the experimental data of charmonium and bottomonium masses. We will show this later.      

In order to solve the three-body Schr\"odinger equation, we first introduce the Jacobi coordinates,
\begin{eqnarray}
{\bf R} &=& {1\over \mu}(\mu_1{\bf r}_1+\mu_2{\bf r}_2+\mu_3{\bf r}_3), \nonumber\\
{\bf x}_1&=&{\sqrt{(\mu_1+\mu_2)\mu_3} \over \mu}\left({\bf r}_3-{\mu_1{\bf r}_1+\mu_2{\bf r_2}\over \mu_1+\mu_2}\right),\nonumber\\
{\bf x}_2&=&\sqrt{{\mu_1 \mu_2\over (\mu_1+\mu_2)\mu}}\left({\bf r}_2-{\bf r}_1\right),
\label{jacobi}
\end{eqnarray}
where $\mu \equiv \sum_{i=1}^3 \mu_i$ is the total mass. With such coordinates, the kinetic energy becomes
\begin{equation}
\sum_{i=1}^3 {\hat {\bf p}^2_i\over 2\mu_i} = \frac{\hat {\bf P}^2+\hat {\bf q}_1^2+\hat {\bf q}_2^2}{2\mu},
\end{equation}
where $\hat {\bf q}_1$ and $\hat {\bf q}_2$ are the relative momenta that conjugate to ${\bf x}_1$ and ${\bf x}_2$, respectively.

Since the potential depends only on the relative coordinates ${\bf x}_i$ and the total momentum is conserved, one can factorize the three-body motion into a center-of-mass motion and a relative motion, $\Psi({\bf r}_1,{\bf r}_2,{\bf r}_3)=e^{i{\bf P \cdot R}}\Phi({\bf x}_1,{\bf x}_2)$. The bound state properties only relate to the relative motion of the system, and we just need to deal with the six-dimensional wave equation. We then express the relative coordinates ${\bf x}_1$ and ${\bf x}_2$ in the hyperspherical frame~\cite{Krivec:1998}: hyperradius $\rho=\sqrt{x_1^2+x_2^2}$ and hyperangles $\Omega=\{\alpha_2, \theta_1,\phi_1, \theta_2,\phi_2 \}$, where the angle $\alpha_2 \equiv \arcsin(x_2/ \rho)$ is defined within the range $[0,\pi/2]$, and $\{x_i, \theta_i, \phi_i\}$ are the spherical coordinates corresponding to ${\bf x}_i$. With the hyperspherical coordinates, the Schr\"odinger equation governing the relative wave function $\Phi(\rho,\Omega)$ can be written as 
\begin{eqnarray}
\left[ {1\over 2\mu}\left( -{d^2 \over d\rho^2}-{5\over \rho}{d \over d\rho}  + {\widehat {\bf K}^2\over \rho^2}\right) + V(\rho, \Omega) \right ]\Phi = E_r \Phi,
\label{relative}
\end{eqnarray}
where the corresponding energy eigenvalue $E_r = E-\frac{{\bf P}^2}{2M}$ and $\widehat {\bf K}$ is the hyperangular momentum operator.
The hyperspherical harmonic(HH) functions ${\mathcal Y}_\kappa(\Omega)$ are the eigenstates of $\widehat {\bf K}^2$,
\begin{equation}
\widehat {\bf K}^2{\mathcal Y}_\kappa(\Omega)=K(K+4){\mathcal Y}_\kappa(\Omega),
\end{equation}
where $K$, referred to as the grand-orbital momentum, is the quantum number describing the magnitude of the hyper-angular momentum. There are eight operators that commute with the kinetic energy term in the Hamiltonian and with each other. In addition to $K$, the other conserved quantum numbers are total angular momentum($L$), total magnetic quantum number($M$), angular momentum of corresponding to each Jacobi coordinate($l_1$ and $l_2$), and $n$. Here $l_1$($l_2$) is the orbital angular momentum quantum number of subsystem 1 and 2 (12 diquark and 3). $n$ is the relative radial quantum number. These quantum numbers satisfy, 
\begin{eqnarray}
K&=&2n+L, \nonumber\\
L&=&l_1+l_2, \nonumber\\
M&=&m_1+m_2.
\end{eqnarray}
We have introduced the shorthand that
$\kappa \equiv \{K, L, M, n, l_1,  l_2\}$.
In a three-body system, the HH function can be expressed as, 
\begin{align}
\begin{split}
{\mathcal Y}_{\kappa}(\Omega)=&
    \sum_{m_1,m_2} \langle l_1m_1l_2m_2|LM\rangle Y_{l_1}^{m_1}(\theta_1,\phi_1)Y_{l_2}^{m_2}(\theta_2,\phi_2)
\\&
    \times  \left[ {\mathcal N}\, \sin^{l_2}\alpha_2 \cos^{l_1}\alpha_2\, P_{n}^{l_2+\frac{1}{2}, l_{1}+\frac{1}{2}}(\cos 2 \alpha_2) \right],
\end{split}
\end{align}
with $Y_{l}^{m}$ being the spherical harmonics, $P_n^{l}$ the associated Legendre polynomials, and
\begin{eqnarray}
&&{\mathcal N}=\sqrt{ (2K+4)n! \Gamma(n+l_{1}+l_2+2) \over \Gamma(n+l_2+{3\over 2}) \Gamma(n+l_{1}+{3\over 2})}. 
\end{eqnarray}
See e.g.,~\cite{Barnea:1999be,Barnea:2006sd,Marcucci:2019hml} for properties of the HH functions. 

\subsection{Spatial, Color, and Spin Wavefunctions}
As shown in \eqref{v2}, the potential $V(\rho, \Omega)$ depends on the color and spin degrees of freedom. We start to construct the color and spin wave-function, based on the symmetry properties.
Identical fermions should fulfill the Pauli principle, the total wave-function should be antisymmetric under exchange. For heavy quarks, the flavor wave function is trivial. So, the total wave function can be expressed as,
 \begin{eqnarray}
\Psi=\psi_\mathrm{space}\,\phi_\mathrm{color}\,\chi_\mathrm{spin},
\end{eqnarray}
where the color wave-function $\phi_\mathrm{color}=(QQ)_{\bar 3_c}Q_{3_c}$ is antisymmetry. For the spin space, one has,
\begin{eqnarray}
2\otimes 2 \otimes 2= (3\oplus 1)\otimes 2= 4 \oplus 2\oplus 2.
\end{eqnarray}
As a result, the othonormal basis is listed as follows,
\begin{eqnarray}
\chi_{+\frac{3}{2}}^s&=&
    \left|\frac{3}{2},+\frac{3}{2}\right>_{\mathrm{S}} =\left| \uparrow \uparrow \uparrow \right>  \nonumber\\
\chi_{+\frac{1}{2}}^s&=&
    \left|\frac{3}{2},+\frac{1}{2}\right>_{\mathrm{S}} = 3^{-1/2} \Big[ \left| \uparrow \uparrow \downarrow \right> + \left| \uparrow \downarrow \uparrow \right> + \left| \downarrow \uparrow \uparrow \right> \Big] \nonumber\\
\chi_{-\frac{1}{2}}^s&=&
    \left|\frac{3}{2},-\frac{1}{2}\right>_{\mathrm{S}} = 3^{-1/2} \Big[ \left|\downarrow \downarrow \uparrow  \right> + \left| \downarrow \uparrow \downarrow \right> + \left| \uparrow \downarrow \downarrow \right> \Big] \nonumber\\
\chi_{-\frac{3}{2}}^s&=&
    \left|\frac{3}{2},-\frac{3}{2}\right>_{\mathrm{S}} =\left| \downarrow \downarrow \downarrow \right> \nonumber\\
\chi_{+\frac{1}{2}}^{ms}&=&
    \left|\frac{1}{2},+\frac{1}{2}\right>_{\mathrm{MS}} = 6^{-1/2} \Big[ \left| \uparrow \downarrow \uparrow \right> +\left| \downarrow \uparrow \uparrow \right>-2\left| \uparrow \uparrow \downarrow \right> \Big] \nonumber\\
\chi_{-\frac{1}{2}}^{ms}&=&
    \left|\frac{1}{2},-\frac{1}{2}\right>_{\mathrm{MS}} = 6^{-1/2} \Big[ 2\left| \downarrow \downarrow \uparrow \right> - \left| \downarrow \uparrow \downarrow\right> - \left|  \uparrow \downarrow\downarrow \right> \Big] \nonumber\\
\chi_{+\frac{1}{2}}^{ma}&=&
    \left|\frac{1}{2},+\frac{1}{2}\right>_{\mathrm{MA}} = 2^{-1/2} \Big[ \left| \uparrow \downarrow \uparrow \right> - \left| \downarrow \uparrow \uparrow \right> \Big] \nonumber\\
\chi_{-\frac{1}{2}}^{ma}&=&
    \left|\frac{1}{2},-\frac{1}{2}\right>_{\mathrm{MA}} = 2^{-1/2} \Big[ \left|  \uparrow \downarrow\downarrow \right>-  \left| \downarrow \uparrow \downarrow \right>  \Big] 
\end{eqnarray}
Here the $ms$ or $ma$ represent exchange antisymmetry or symmetry between 1 and 2 particle.
It is easy to get,
\begin{eqnarray}
{\bf s}_i\cdot {\bf s}_j |\chi^s\rangle&=&{1\over 4} |\chi^s\rangle,  \nonumber\\
{\bf s}_1\cdot {\bf s}_2 |\chi^{ms}\rangle&=&{1\over 4} |\chi^{ms}\rangle, \nonumber\\
{\bf s}_1\cdot {\bf s}_3 |\chi^{ms}\rangle&=& -{1\over 2}|\chi^{ms}\rangle-{\sqrt{3}\over 4}|\chi^{ma}\rangle, \nonumber\\
{\bf s}_2\cdot {\bf s}_3 |\chi^{ms}\rangle&=& -{1\over 2}|\chi^{ms}\rangle+{\sqrt{3}\over 4}|\chi^{ma}\rangle, \nonumber\\
{\bf s}_1\cdot {\bf s}_2 |\chi^{ma}\rangle&=&-{3\over 4} |\chi^{ma}\rangle, \nonumber\\
{\bf s}_1\cdot {\bf s}_3 |\chi^{ma}\rangle&=& -{\sqrt{3}\over 4}|\chi^{ms}\rangle, \nonumber\\
{\bf s}_2\cdot {\bf s}_3 |\chi^{ma}\rangle&=& {\sqrt{3}\over 4}|\chi^{ms}\rangle,
\end{eqnarray}
where the subscript has been omitted.


Noting that the ground state spacial wavefunction is symmetric, we focus on $S$-wave states (with $L=0$) in this work.
For $\Omega_{ccc}$ and $\Omega_{bbb}$, the spin wavefunction must be symmetric. Therefore, only the $|\phi_c\chi^s\rangle$ state with $J^P={3\over 2}^+$ will be considered. The potential elements in color-spin space can be expressed as,
\begin{align}
\begin{split}
&\langle\phi_c\chi^s|\sum_{i<j}V_{ij}|\phi_c\chi^s\rangle \\
=\;& \frac{V^c_{12}+ V^c_{13}+V^c_{23}}{2}
+ \frac{V^{ss}_{12}+V^{ss}_{13}+V^{ss}_{23}}{8}.
\end{split}
\label{ccc}
\end{align}
For the $\Omega_{ccb}$ and $\Omega_{bbc}$, the spin wavefunction should be symmetric when exchanging the first two particles. The ground states can be $|\phi_c\chi^s\rangle$ with $J^P={3\over 2}^+$or $|\phi_c\chi^{ms}\rangle$ with $J^P={1\over 2}^+$. The potential elements of the former are given in Eq.~\eqref{ccc}, whereas those of the latter can be written as,
\begin{align}
\begin{split}
&\langle\phi_c\chi^{ms}|\sum_{i<j}V_{ij}|\phi_c\chi^{ms}\rangle\\
=\;&\frac{V^c_{12}+V^c_{13}+V^c_{23}}{2}
+\frac{V^{ss}_{12}}{8}-\frac{V^{ss}_{13}+V^{ss}_{23}}{4}.
\end{split}
\end{align}

\subsection{Hyperspherical Harmonics Expansion}
The potential $V(\rho, \Omega)$ depends not only on the hyperradius but also on the eight hyperangles. The Schr\"odinger equation~\eqref{relative} cannot be further factorized into a radial part and an angular part. Instead, one expands the wave function in terms of the hyper-spherical harmonic(HH) functions ${\mathcal Y}_\kappa(\Omega)$. The total relative wave function can be expanded as
 \begin{eqnarray}
  \Phi(\rho, \Omega)=\sum_\kappa R_\kappa(\rho){\mathcal Y}_\kappa(\Omega)|\phi_c\chi^s\rangle,
  \end{eqnarray}
for the ${3\over 2}^+$ states, and
 \begin{eqnarray}
 \Phi(\rho, \Omega)=\sum_\kappa R_{\kappa}(\rho){\mathcal Y}_\kappa(\Omega) |\phi_c\chi^{ms}\rangle\,,
\end{eqnarray}
for the ${1\over 2}^+$ states.
Taking such an expansion and employing the reduced radial wavefunction as $u_\kappa(\rho) \equiv \rho^{5/2} R_\kappa(\rho)$, the multi-variable Schr\"oedinger equation becomes a set of coupled differential equations of single variable,
\begin{eqnarray} 
\left[{1\over 2\mu}{d^2 \over d\rho^2} - {4K(K+4) + 15 \over 8\mu \,\rho^2}+E_r \right]u_{\kappa}=\sum_{\kappa'}V_{\kappa \kappa'}u_{\kappa'},
\label{eq.cde}
\end{eqnarray}
where $V_{\kappa \kappa'}$ is the potential matrix,
\begin{align}
\begin{split}
V_{\kappa \kappa'} 
=\;&\int {\mathcal Y}_\kappa^*(\Omega)V(\rho, \Omega){\mathcal Y}_{\kappa'}(\Omega)d\Omega \\
=\;&\sum_{i<j}\int V_{ij}(|\mathbf{r}_{ij}|){\mathcal Y}_\kappa^*(\Omega){\mathcal Y}_{\kappa'}(\Omega)\mathrm{d}\Omega\,,
\end{split}\label{vmatrix}
\end{align}
with the volume element being
\begin{equation}
d\Omega= \cos^2 \alpha_2 \sin^2 \alpha_2 \sin \theta_1 \sin \theta_2 d\alpha_2d\theta_1d\phi_1d\theta_2d\phi_2.
\end{equation}

\subsection{Computation of the Potential Matrix}
Computing the potential matrix is nontrivial. In the most general form, \eqref{vmatrix} is a five-dimensional integral, which is computationally expensive. However, taking the assumption that the total interaction potential is the summation of two-body interaction $V_{ij}(|{\bf r}_{ij}|)$, one can reduce \eqref{vmatrix} into a one-dimensional integral by performing particle permutation.
Let us first focus on $V_{12}$, which depends only on ${\bf x}_2$, see Eq.~\eqref{jacobi}. The integral over $\alpha_2$ is the only non-trivial one out of the five hyperangles, whereas the remainders can be computed analytically using the orthogonal relation of the HH functions.
To compute other sectors of the potential matrix, the trick is to consider another form of Jacobi coordinate, which differs from the previous one by a particle permutation and puts ${\bf r}_2-{\bf r}_3$ or ${\bf r}_3 - {\bf r}_1$ in the newly defined ${\bf x}_2$. We label the covention in Eq.~\eqref{jacobi} as ${\bf x}^{(12)}_{i}$, and the other two conventions as ${\bf x}^{(23)}_{i}$ and ${\bf x}^{(31)}_{i}$.
Explicitly, they are defined as
\begin{eqnarray}
{\bf x}^{(23)}_1&=&{\sqrt{(\mu_2+\mu_3)\mu_1} \over \mu}\left({\bf r}_1-{\mu_2{\bf r}_2+\mu_3{\bf r_3}\over \mu_2+\mu_3}\right),\nonumber\\
{\bf x}^{(23)}_2&=&\sqrt{{\mu_2 \mu_3\over (\mu_2+\mu_3)\mu}}\left({\bf r}_3-{\bf r}_2\right),
\label{jacobi2}
\end{eqnarray}
and
\begin{eqnarray}
{\bf x}^{(31)}_1&=&{\sqrt{(\mu_3+\mu_1)\mu_2} \over \mu}\left({\bf r}_2-{\mu_3{\bf r}_3+\mu_1{\bf r_1}\over \mu_3+\mu_1}\right),\nonumber\\
{\bf x}^{(31)}_2&=&\sqrt{{\mu_3 \mu_1\over (\mu_3+\mu_1)\mu}}\left({\bf r}_1-{\bf r}_3\right),
\label{jacobi3}
\end{eqnarray}
which can be related to ${\bf x}^{(12)}$ by
\begin{align}
\left(\begin{array}{c}
{\bf x}_1^{(12)} \\
{\bf x}_2^{(12)} \\
\end{array}\right)
= \left(\begin{array}{cc}
a_{11}& a_{12} \\
a_{21}& a_{22} \\
\end{array}\right)\cdot
\left(\begin{array}{c}
{\bf x}_1^{(ij)} \\
{\bf x}_2^{(ij)} \\
\end{array}\right),
\end{align}
where $a_{ij}$ is element of the rotation matrix. 

It has been shown that under the particle permutation, different HH's are related by an unitary transformation,
\begin{eqnarray}
{\mathcal Y}_{\kappa}(\Omega) = \sum_{\kappa'} R^{(ij)}_{\kappa \kappa'} {\mathcal Y}_{\kappa'}(\Omega^{(ij)}),
\end{eqnarray}
and the coefficients $R^{(ij)}_{\kappa \kappa'}$ are refereed to as the Raynal--Revai coefficients~\cite{Raynal:1970ah}. 
The Raynal--Revai coefficient is non vanishing only for the HH's with the same grand-orbital momentum($K$), total angular momentum($L$), and magnetic quantum number($M$).
Details of the Raynal--Revai coefficients, especially its explicit form for $K=0,2,4,6$ are shown in Appendix~\ref{sec.appendix}.

With these preparation, we are now ready to compute the potential matrix for an arbitrary $(i,j)$ pair:
\begin{eqnarray}
V_{\kappa\kappa'}^{(ij)} (\rho) &\equiv&\int V^{(ij)} (|\mathbf{r}_{ij}|) {\mathcal Y}_\kappa^*(\Omega) {\mathcal Y}_{\kappa'}(\Omega)\mathrm{d}\Omega \nonumber\\
&=& \sum_{kk'} (R^{(ij)}_{\kappa k})^{*} R^{(ij)}_{\kappa' k'} 
	\nonumber\\
&\times &
    \int V^{(ij)} (|\mathbf{r}_{ij}|)
	{\mathcal Y}_k^*(\Omega^{(ij)}) {\mathcal Y}_{k'}(\Omega^{(ij)})\mathrm{d}\Omega^{(ij)} \nonumber\\
&=&\sum_{kk'} (R^{(ij)}_{\kappa k})^{*} R^{(ij)}_{\kappa' k'} V_{kk'}^{(12)}(\rho).
\end{eqnarray}

\section{Baryons in vacuum}
\label{section3}

\begin{table}[!bt]
\renewcommand\arraystretch{1.5}
\caption{Potential model parameters}
\label{tab1}
\setlength{\tabcolsep}{2.5mm}
\begin{tabular}{cccc}
	\toprule[1pt]\toprule[1pt] 
	$m_b$ & $m_c$ & $\alpha$ & $\sigma$ \\
	4.7~GeV & 1.29~GeV & 0.308 & 0.15~GeV$^2$ \\
	\midrule[0.7pt]
    $\gamma$ & $\beta_{b\bar b}$ & $\beta_{c\bar c}$ & $\beta_{b\bar c}$=$\beta_{c\bar b}$\\
    1.982~GeV & 0.239~GeV & 1.545~GeV & 0.525~GeV
	\\
	\bottomrule[1pt]\bottomrule[1pt]
\end{tabular}
~\\~\\
\caption{The experimental~\cite{10.1093/ptep/ptaa104} and calculated quarkonium masses}
\label{tab2}
\setlength{\tabcolsep}{2.5mm}
\begin{tabular}{c|cc}
\toprule[1pt]\toprule[1pt] 
 State & $M_{\rm exp}$(GeV) & $M_{\rm theo}$(GeV)\\
\midrule[0.7pt]
 $\eta_c$ 	    & 2.981 & 2.968 \\
 $J/\psi$ 	    & 3.097 & 3.102 \\
 $h_c(1P)$ 	    & 3.525 & 3.480 \\
 $\chi_c(1P)$ 	& 3.556 & 3.500 \\
 $\eta_c(2S)$ 	& 3.639 & 3.654 \\
 $\psi(2S)$ 	& 3.696 & 3.720 \\
 $\eta_b$       & 9.398 &  9.397 \\
 $\Upsilon(1S)$ & 9.460 &  9.459 \\
 $h_b(1P)$      & 9.898 &  9.845 \\
 $\chi_b(1P)$   & 9.912 &  9.860 \\
 $\eta_b(2S)$   & 9.999 &  9.957 \\
 $\Upsilon(2S)$ & 10.023 &  9.977 \\
 $B_c(^1S_0)$ 	&	6.275 &  6.282 \\
 $B_c(^3S_1)$ 	&	- &  6.347 \\
 $B_c(^1P_1)$	&	- &  6.726 \\
 $B_c(^3P_{0,1,2})$ 	&	- &  6.738 \\
 $B_c(2^1S_0)$ 	&	6.871 &  6.886 \\
 $B_c(2^3S_1)$	&	- &  6.915 \\
\bottomrule[1pt]\bottomrule[1pt]
\end{tabular}
\end{table}

\begin{table*}[!hptb]
\renewcommand\arraystretch{2.0}
\caption{The calculated fully heavy flavor baryon mass $M_{\rm theo}$ and the root-mean-squared radius $r_{\text{rms}}$ for the ground $1S$ and radial-excited states, $2S$ and $3S$.}
\label{tab3}\setlength{\tabcolsep}{1.2mm}
\begin{tabular}{c|ccc|ccc|ccc|ccc|ccc|ccc}
\toprule[1pt]\toprule[1pt] 
    &\multicolumn{12}{c|}{$J^{P}={3\over 2}^+$}
    &\multicolumn{6}{c}{$J^{P}={1\over 2}^+$} \\
    & \multicolumn{3}{c}{$\Omega_{ccc}$} & \multicolumn{3}{c}{$\Omega_{ccb}$} 
    & \multicolumn{3}{c}{$\Omega_{bbc}$} & \multicolumn{3}{c|}{$\Omega_{bbb}$} 
    & \multicolumn{3}{c}{$\Omega_{ccb}$} & \multicolumn{3}{c}{$\Omega_{bbc}$}  \\
\midrule[0.7pt]
    \multicolumn{1}{c|}{State}&  
    $1S$ & $2S$ & $3S$ & $1S$ & $2S$ & $3S$ & 
    $1S$ & $2S$ & $3S$ & $1S$ & $2S$ & $3S$ & 
    $1S$ & $2S$ & $3S$ & $1S$ & $2S$ & $3S$ \\
\midrule[0.7pt]
\multirow{1}{*}
{$M_{\rm theo}$(GeV)} 
    & 4.80  & 5.31  & 5.74  
    & 8.19  & 8.68  & 9.08 
    & 10.96 & 11.33 & 11.63 
    & 14.36 & 14.77 & 15.09 
    & 8.17  &  8.66 & 9.07  
    & 10.87 & 11.28 & 11.59 \\
$r_{\text{rms}} $(fm) 
    & 0.30 & 0.45 & 0.57 
    & 0.22 & 0.33 & 0.43 
    & 0.19 & 0.31 & 0.41 
    & 0.16 & 0.26 & 0.34
    & 0.22 & 0.33 & 0.42  
    & 0.18 & 0.30 & 0.40
\tabularnewline
\bottomrule[1pt]\bottomrule[1pt]
\end{tabular}
\end{table*}
\begin{figure*}[!htb]
    \includegraphics[width=0.99\textwidth]{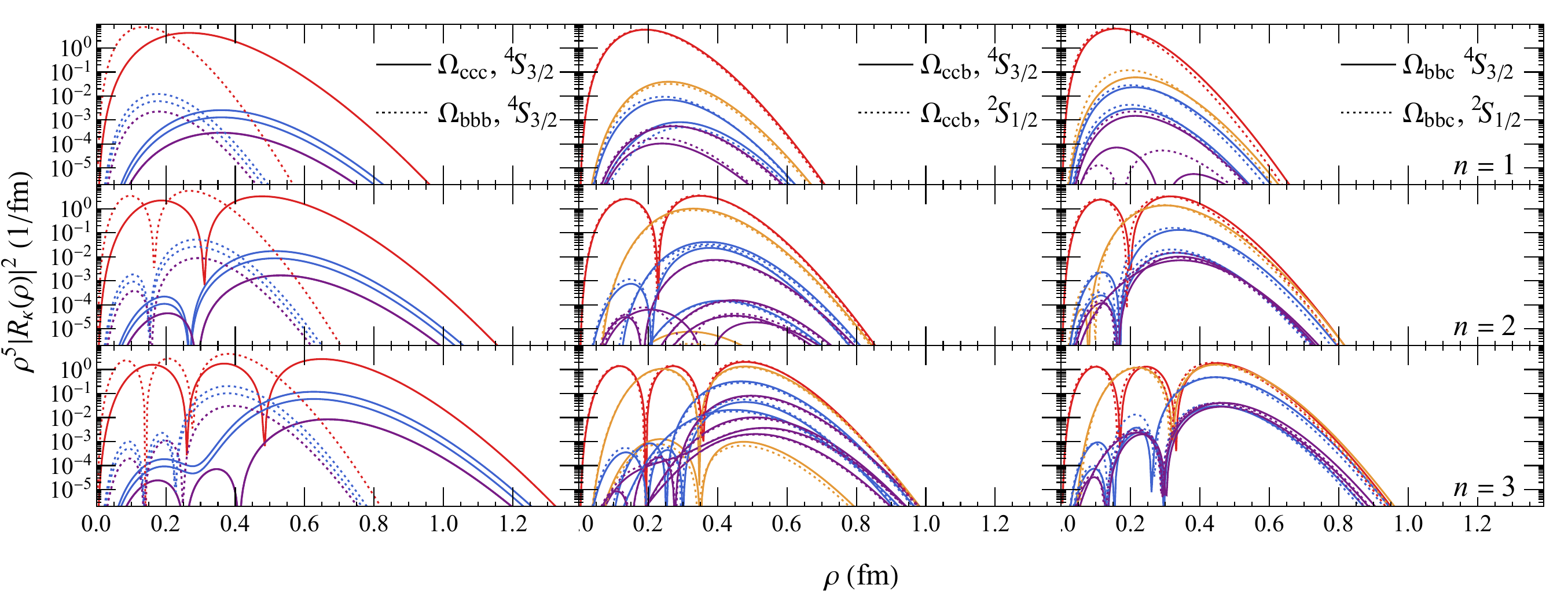}
    \caption{(Left) The vacuum radial probabilities for $1S$(top), $2S$(mid), and $3S$(bottom) states of $\Omega_{ccc}$(solid) and $\Omega_{bbb}$(dotted) with quantum number $J^{P}=(3/2)^{+}$. 
    The same for $\Omega_{ccb}$(middle) and $\Omega_{bbc}$(right) particles with $J^{P}=(3/2)^{+}$(solid) and $J^{P}=(1/2)^{+}$(dotted).
    Lines with different colors correspond to $K=0$(red), $2$(gold), $4$(blue), and $6$(purple). See texts for detailed explanation of the states.
	\label{fig.wf_vacuum}}
\end{figure*}

In this section, we start with computing the heavy flavor baryons $QQQ$ bound states in a vacuum. 
As mentioned before, we first fix the parameters in potential and the quark masses ($m_c$ and $m_b$) with the knew charmonium, bottomonium, and also $B_c$ mesons data.
Their masses can be calculated via the two-body Schr\"odinger equation with the potential
\begin{equation}
V_{Q\bar Q} = {4\over 3}\left(V^c_{ij}(r) + V^{ss}_{ij}(r){\bf s}_i\cdot {\bf s}_j\right),  
\end{equation}
where $4/3$ is the color factor for color-singlet states $Q\bar Q$. With the model parameters presented in Table~\ref{tab1}, we obtain the quarkonium masses and $B_c$ mesons shown in Table~\ref{tab2}. Here the parameters and quarkonium masses have been used and calculated in previous work~\cite{Zhao:2020nwy}. One can see that, the quarkonia mass can be well described.

With the known parameters, we then solve the three-body Schr\"odinger equations \eqref{eq.cde} for fully-heavy baryons $QQQ$.
The baryon mass comes from the summation of the constituent masses $\mu =\sum_{i=1}^{3} \mu_i$ and the binding energy $E_r$ which is determined by the radial equations,
\begin{equation}
M_B = \mu+E_r \,,
\end{equation}
and the root-mean-squared radius is defined as
\begin{eqnarray}
r_{\text{rms}}^2 = \int \sum_\kappa|R_\kappa(\rho)|^2 \rho^{7} d\rho.
\label{rms}
\end{eqnarray}
One may find $r_{\text{rms}}^2 = \big\langle {1\over 3}\sum_{i=1}^3 ({\bf r}_i-{\bf X})^2\big\rangle$ when all the three quarks have the same mass. The normalization $\int \sum_\kappa|R_\kappa(\rho)|^2 \rho^5 d\rho=1$ for the radial functions $R_\kappa(\rho)$. 

One can only include a finite number of hyperspherical harmonics in a numerical calculation, and our truncation is made according to the symmetry properties of the system. Since we focus in this work on the $S$-wave baryon states, the relevant hyperspherical harmonics are those corresponding to vanishing total orbital angular momentum $L$ and magnetic quantum number $M$, i.e. $L=M=0$. We choose all such hyperspherical harmonic functions with hyperangular quantum number $K\leq 6$.
This leads to coupled differential equations which are numerically solved by using the inverse power method~\cite{CRATER1994470}. The main advantage of taking the inverse power method is its high precision for both ground and excited states.

We show the baryon wavefunctions for the ground state $1S$, first radial excited state $2S$, and the second radial excited state $3S$ states of $\Omega_{ccc}$, $\Omega_{ccb}$, $\Omega_{bbc}$, and $\Omega_{bbb}$ in Fig.~\ref{fig.wf_vacuum} and present the mass and root-mean-squared radius in Table~\ref{tab3}.
There are respectively one, two, three, and four states with quantum number $K=0$, $2$, $4$, and $6$. They are respectively represented by red, gold, blue, and purple curves in Fig.~\ref{fig.wf_vacuum}. The clear hierarchy of their magnitudes shows the convergence of the hyperspherical harmonic expansion. For the equal-mass system, i.e., $\Omega_{ccc}$ and $\Omega_{bbb}$, we notice the vanishment of the $K=2$ states, which always correspond to $n=1$. When interchanging the coordinates of two quarks, $S$ states of the equal-mass system is always symmetric whereas $n=1$ states are anti-symmetric. Thus, $\Omega_{ccc}$ and $\Omega_{bbb}$ do not contain $K=2$ component.
For the unequal-mass baryons, i.e., $\Omega_{ccb}$ and $\Omega_{bbc}$, the $K=2$ states become non-vanishing, and the contributions of other higher order hyperspherical harmonic functions are more sizable. Comparing the spin half states of $\Omega_{ccb}$ and $\Omega_{bbc}$ are respectively lighter than the corresponding three-halves states, we find the spin half states are, respectively, lighter than the corresponding three-halves state. The modest differences in the mass difference and the wave functions indicate the spin-spin interaction is a higher-order effect compared to the central interaction.
We also check the convergence of hyperspherical harmonics expansion by observing same results, up to desired accuracy, when keeping states with $K\leq 4$ only. The ground state masses of these fully heavy baryons are consistent with previous studies based on lattice QCD~\cite{Padmanath:2013zfa, PACS-CS:2013vie,  Meinel:2012qz, Briceno:2012wt}, the potential model~\cite{Silvestre-Brac:1996myf, Martynenko:2007je, Vijande:2015faa, Liu:2019vtx, Yang:2019lsg, Shi:2019tji}, and other effective theories~\cite{Hasenfratz:1980ka, Zhang:2009re, Wang:2011ae, Migura:2006ep}.

\section{Baryons in hot medium}
\label{section4}
\begin{figure}[!htb]
	\includegraphics[width=0.4\textwidth]{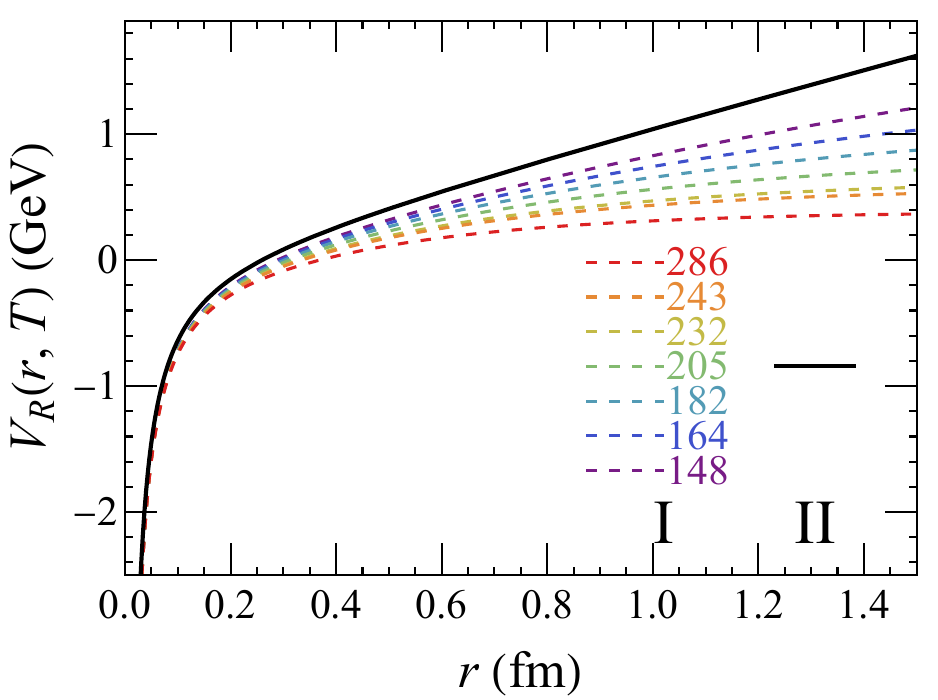} 
	\includegraphics[width=0.4\textwidth]{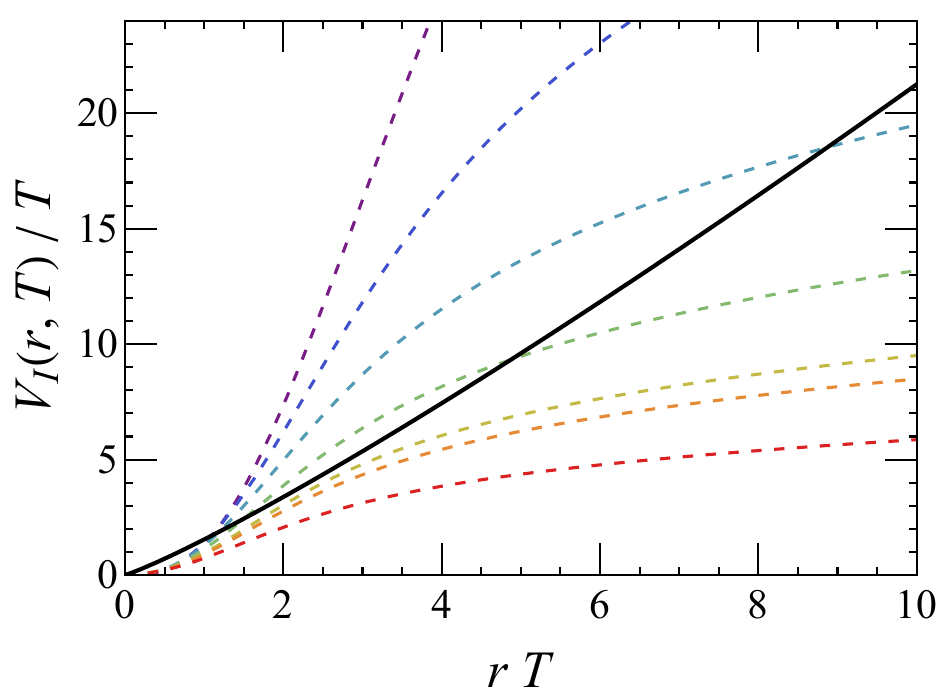} 
	\caption{Distance dependence of real(upper) and temperature-scaled imaginary(lower) potentials taking potential I~\protect{\cite{Burnier:2014ssa,Burnier:2015tda}}(colored dashed) and II~\protect{\cite{Bala:2021fkm}}(black solid). Results for Potential II are temperature independent.}
	\label{fig.V}
\end{figure}

As mentioned before, a hot QCD medium---quark-gluon plasma(QGP) is created in relativistic heavy ion collisions.
The typical temperature of the QGP is hundreds of MeV, estimated by the spectrum of the direct photon~\cite{ALICE:2022wpn}, which is much larger than the binding energy of most hadrons. Only the tightly bound states of heavy quarks, such as $J/\psi$ and $\Upsilon$, can survive in the QGP but with a large thermal width, as shown in lattice results~\cite{Lafferty:2019jpr, Larsen:2019bwy, Larsen:2019zqv}. 
In this section, we come to study the finite-temperature properties of fully heavy baryon states.

\begin{figure*}[!htb]
	$$\includegraphics[width=0.25\textwidth]{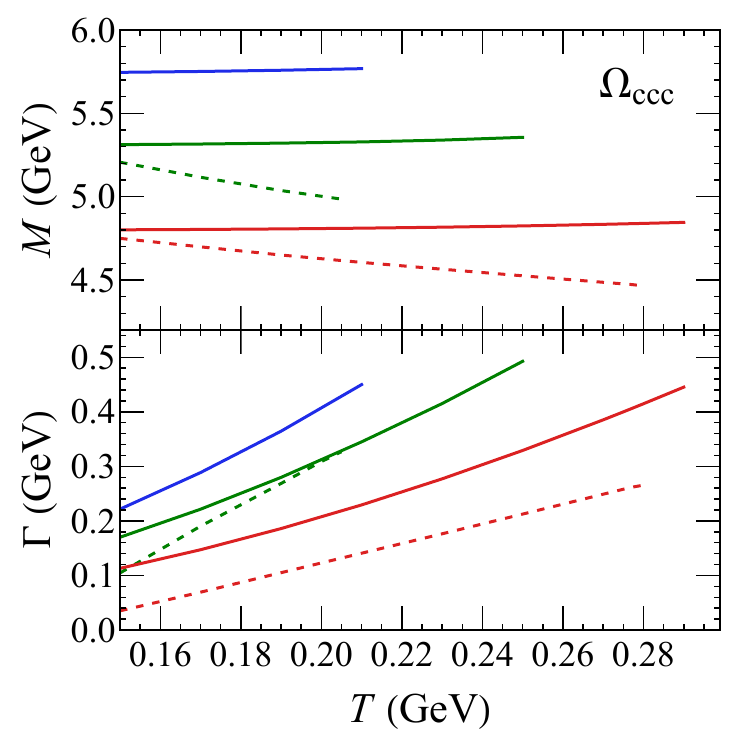}
    \includegraphics[width=0.25\textwidth]{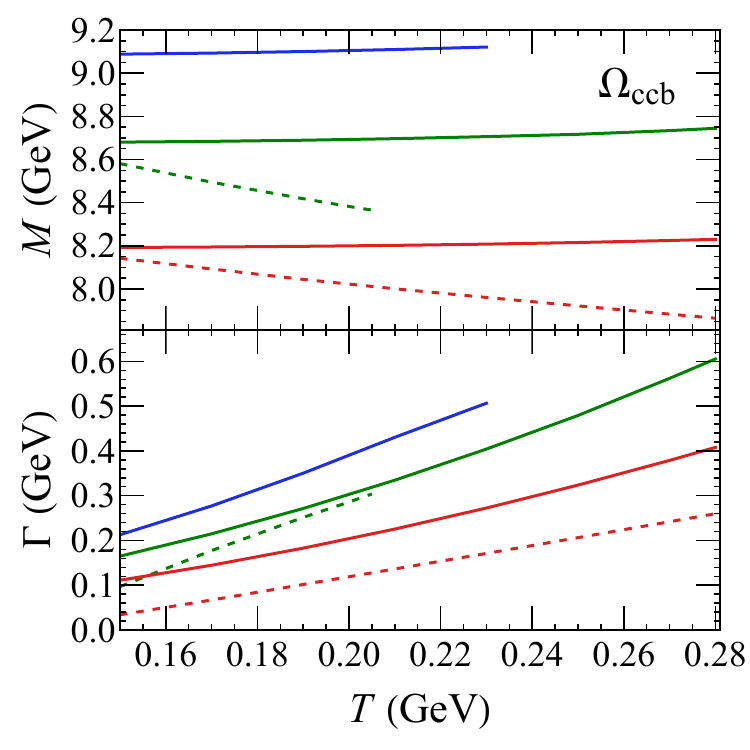}
    \includegraphics[width=0.25\textwidth]{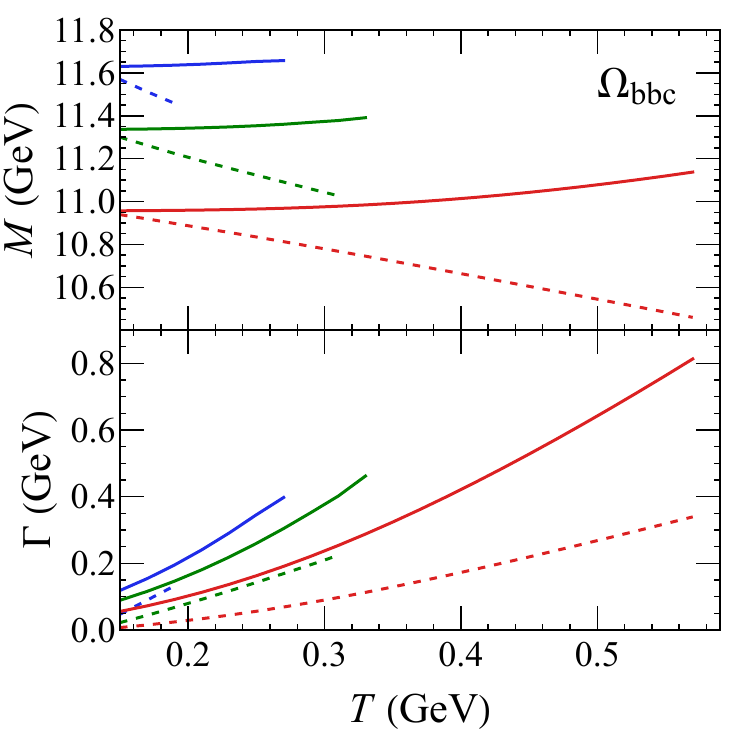}
    \includegraphics[width=0.25\textwidth]{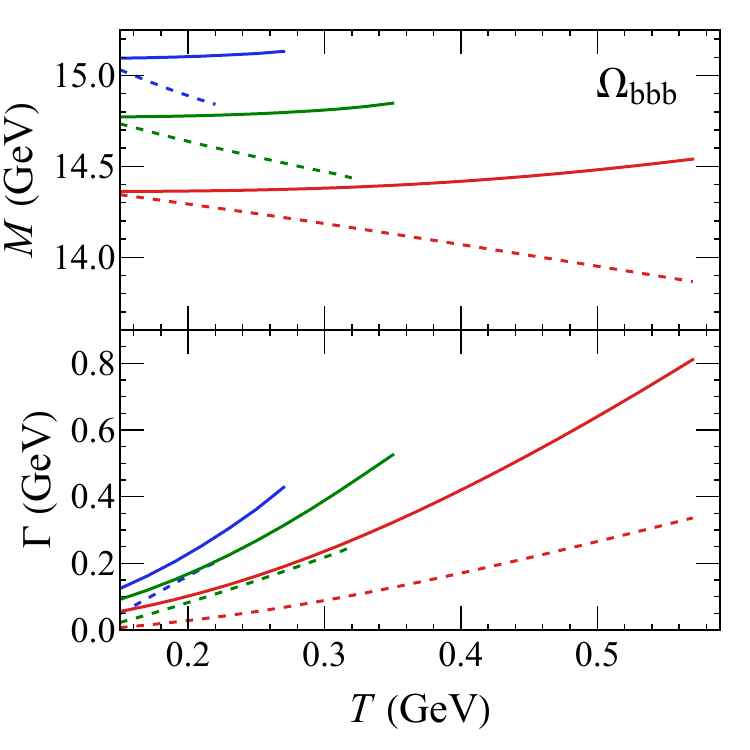}$$
	\caption{(Left to right) Masses(upper) and thermal widths(lower) for $1S$(red), $2S$(green), and $3S$(blue) states of $\Omega_{ccc}$, $\Omega_{ccb}$, $\Omega_{bbc}$, and $\Omega_{bbb}$ as functions of temperature $T$. The solid lines are with potential II~\protect{\cite{Bala:2021fkm}}, while the dashed lines are with potential I~\protect{\cite{Burnier:2014ssa,Burnier:2015tda}}.}
	\label{fig.finiteT}
\end{figure*}
The finite temperature properties of quarkonium states are encoded in the finite temperature potential between heavy quark $Q$ and antiquark $\bar Q$. 
For the baryons, there are no study on their finite-temperature potential in either weak or strong coupling regions. In the weak coupling limit, the HTL study shows the heavy quark potential is proportional to the color factor~\cite{Laine:2006ns}. So, for the color wave-function $(QQ)_{\bar 3_c}Q_{3_c}$ of the baryons, both the real and imaginary potential still satisfy the same relation as shown in Eq.~\eqref{v2}. Besides, we neglect the influence of hot QCD medium to the heavy quark spin-spin interaction.

The heavy-quark interaction potential is screened by other color objects in the QGP, and consequently the long-range interaction (\ref{eq.cornell}) is strongly suppressed when the temperature is high enough, as shown in the hard-thermal-loop (HTL) perturbative calculation~\cite{Laine:2006ns}. Besides, the potential develops an imaginary part which originates from the Landau-damping~\cite{Laine:2006ns}.
While in the strong coupling region, the heavy quark potential can be extracted from the Wilson loop in lattice QCD~\cite{Burnier:2014ssa, Burnier:2015tda, Bala:2021fkm}. The result also shows the heavy quark potential is complex-valued. 
The real part potential shows the screening effect, while the imaginary part reflect the decay of quarkonium under scattering with thermal partons. However, the value of the potential depends on the extraction strategies, the detail as shown in~\cite{Bala:2021fkm}.
To account for the theoretical uncertainty, we choose two qualitatively different schemes of finite-temperature potentials, both of which are given by recent lattice QCD calculation, in this study. 

The first potential scheme is the from Ref.~\cite{Burnier:2015tda}, which has a obvious color-screened real potential, named potential (I). Both the real and imaginary potentials can be fitted by a functional form based on the Gauss' law approach~\cite{Lafferty:2019jpr}, 
\begin{eqnarray}
{\rm Re} V(T,r)&=&-\alpha\left( m_D+{e^{-m_D r}\over r}\right)\\
&+&{\sigma \over m_D}\left(2-(2+m_Dr)e^{-m_Dr}\right), \nonumber\\
{\rm Im}V(T,r)&=&-\alpha T\phi(m_Dr)\nonumber\\
&-&{\sqrt{\pi} \over 4}m_D T \sigma r^3 G_{2,4}^{2,2}\left(_{{1\over2},{1\over2},-{3\over2},-1 }^{-{1\over2},-{1\over2}} \Big| {1\over 4}m_D^2r^2 \right)\nonumber
\end{eqnarray}
with
\begin{eqnarray}
\phi(x)=2\int_0^\infty dz{z\over (z^2+1)^2}\left( 1-{\sin(xz)\over xz} \right),
\end{eqnarray}
where $G$ is the Meijer-G function. When going into a high-temperature region, ${\rm Im}V(r)$ is consistent with the result from the pure HTL theory~\cite{Laine:2006ns}. Both ${\rm Re}V$ and ${\rm Im}V$ only depend on a single temperature dependent parameter, the Debye mass $m_D(T)$, which can be obtained by fitting the lattice data~\cite{Burnier:2014ssa,Burnier:2015tda}. $\alpha$ and $\beta$ in the potential is the same as the vacuum case. 

The second potential scheme is from Ref.~\cite{Bala:2021fkm}, which is the result of Lattice QCD with dynamical fermions and extracted by assuming Gaussian spectral function. We call it potential (II). In this case, the real part potential exhibits negligible screening effect even up to $\sim700$~MeV, and it can be parameterized via Eq.~\eqref{eq.cornell}. Meanwhile, the imaginary part is found to follow a simple form that $V_I/T$ is a single-variable function of $r\,T$. For the relevant region, it value is greater that the potential (I). Potential (II) is quantitatively consistent with the deep-learning extraction of heavy quark potential~\cite{Shi:2021qri} from lattice QCD results of masses and widths~\cite{Larsen:2019bwy, Larsen:2019zqv}.  
The temperature-dependent real and imaginary parts of the potential (I) and (II) are shown by colored and black curves, respetively, in Fig.~\ref{fig.V}.

With the complex potential, we solve the coupled radial equations~\eqref{eq.cde} and both the energy eigenvalues and wavefunctions are also complex-valued. The real part of the energy eigenvalue gives the baryon mass: $M_B(T)=\mu+{\rm Re}[E_r(T)]$, while the imaginary part gives the thermal width of the baryons $\Gamma(T)=-{\rm Im}[E_r(T)]$. The results for fully heavy baryons, $\Omega_{ccc}$, $\Omega_{ccb}$, $\Omega_{bbc}$, and $\Omega_{bbb}$, are shown in Fig.~\ref{fig.finiteT}. First, we can see the large difference in masses and widths between two potential schemes. With potential (I), masses decrease with the temperature while the thermal widths are generated with temperature increases. 
With potential (II), we observe weaker temperature-dependence of masses, owning to fact that the mass is mostly controlled by the real part potential.
The thermal widths are increasing with temperature and are quantitatively larger than the those using potential (I). 
For the $\Omega_{ccc}$ and $\Omega_{ccb}$ the second radial excited states $3S$ disappear when the temperature high than $0.15$ GeV as presented when using the potential (I). Second, we can see that replace one charm to the bottom quark, the thermal decay width of $\Omega_{ccb}$ is very close to $\Omega_{ccc}$ (also for $\Omega_{bbc}$ and $\Omega_{bbb}$). Comparing to the widths of quarkonium with potential (I) computed in Ref.~\cite{Lafferty:2019jpr}, we find the thermal width of $\Omega_{ccc}$ is obviously larger than $J/\psi$, while $\Omega_{bbb}$ is almost same as the $\Upsilon$. 
The results indicate most of the fully-heavy baryons can also survive in a hot QCD medium for a long time in both cases. Meanwhile, with potential (I), there exists an obvious dissociation temperature $T_d$. When the medium temperature is higher than $T_d$, the fully-heavy baryon will disappear immediately. With the potential (II), the fully-heavy baryon can still survive at high temperatures as long as the temperature of the medium drops down fast. These two different behaviors will be reflected in the transport and final production of fully-heavy baryons in relativistic heavy ion collisions. 
Precisely studying the yield of these baryons in the QGP should go beyond the hadronization at phase transition hypersurface as in previous studies~\cite{He:2014tga, Zhao:2016ccp, Cho:2019syk, Andronic:2021erx}.

\section{Summary}
\label{section5}
In this work, we study the properties of fully-heavy baryons in the vacuum and the finite temperature, which is created in high-energy nuclear collisions. 
We employ the Hyperspherical Harmonics expansion and solve the three-body Schr\"odinger equation. 
We obtain the masses and wave functions of $\Omega_{ccc}$, $\Omega_{ccb}$, $\Omega_{bbc}$, and $\Omega_{bbb}$, up to second radial excited states, for both zero and finite temperatures. 
In the vacuum, the ground state masses of predicted are consistent with other studies. 
In a hot medium, the temperature dependence of the baryon masses and the thermal decay widths are calculated. 
These properties are important to their productions in heavy ion collisions, as they are expected to affect the yield and momentum distributions of heavy-flavor hadrons in high-energy nuclear collisions. 

\vspace{5mm}
\noindent {\bf Acknowledgement}: The work is supported by the European Union’s Horizon 2020 research and the innovation program under grant agreement No. 824093 (STRONG-2020) (J.Z.) and Tsinghua University under grant No. 53330500923 (S.S.).

\begin{appendix}
\section{Raynal--Revai coefficient}\label{sec.appendix}

The Raynal--Revai coefficient is non vanishing only for the HH's with the same grand-orbital momentum($K$), total angular momentum($L$), and magnetic quantum number($M$).
A general expression for the Raynal–Revai coefficients is obtained in the form of a multiple sums over the powers of the coefficients $a_{ij}$,
\begin{eqnarray}
R^{(ij)}_{\kappa \kappa'}&\equiv& \langle l_i', l_j'| l_i,l_j\rangle_{K,L} \\
&=&(-1)^{n+n'}(C_{l_i,l_j}^nC_{l_i',l_j'}^{n'})^{-1/2}\sum_{l_1,l_2,l_3,l_4}i^{l_2-l_1+l_j-l_j'}\nonumber\\
&\times&f( l_1, l_3; l_i')f(l_2,l_3; l_i)f(l_2, l_4; l_j')f( l_1, l_4; l_j)\nonumber\\
&\times& {\rm sgn}(a_{12})^{l_1}{\rm sgn}(a_{21})^{l_2}{\rm sgn}(a_{11})^{l_3}{\rm sgn}(a_{22})^{l_4}\nonumber\\
&\times&
\left(
\begin{array}{ccc}
l_3 & l_1 & l_i' \\
l_2 & l_4 & l_j' \\
l_x & l_y & L 
\end{array}
\right) \sum_{\nu_1,\nu_2} (-1)^{\nu_1} C_{l_1l_2}^{\nu_1} C_{l_3l_4}^{\nu_2}\nonumber\\
&\times&  |a_{12}|^{2\nu_1+l_1+l_2} |a_{11}|^{2\nu_2+l_3+l_4}, \nonumber
\end{eqnarray}
where
\begin{eqnarray}
&&C_{jl}^n  \nonumber\\
&&={(2n+j+l+1)!\over n! (n+j+l+1)! (2(n+j)+1)!!(2(n+l)+1)!!},\nonumber\\
\end{eqnarray}
and
 \begin{eqnarray}
f(a,b;c)=\sqrt{(2a+1)(2b+1)}\langle a0 b0 | c 0 \rangle.
\end{eqnarray}
The notation in the three-by-three bracket is the 9j Clebsch--Gordan coefficient. 
The summation is restricted by,
\begin{eqnarray}
K&=&2n+l_i+l_j=2n'+l_i'+l_j'\nonumber\\
&=&2(\nu_1+\nu_2)+l_1+l_2+l_3+l_4.
\end{eqnarray}

For the quantum number of $K=0$ and $L=0$, the Raynal--Revai coefficient $R^{(ij)}=1$.
While for the quantum number of $K=2$ and $L=0$, the Raynal--Revai coefficient is a two-by-two matrix and can be expressed as (here we already take the equalities $a_{11}=a_{22}$ and $a_{12}=-a_{21}$, which is given by the rotation matrix.),
\begin{align}
R^{(ij)}_{2\times 2} = \left(
\begin{array}{cc}
a_{11}^2-a_{12}^2 & -2a_{11}a_{12} \\
2a_{11}a_{12} & a_{11}^2-a_{12}^2
\end{array}
\right).
\end{align}
For the quantum number of $K=4$ and $L=0$, 
\begin{align}
R^{(ij)}_{3\times 3} = \left(
\begin{array}{ccc}
c_{11} & c_{12} & c_{13} \\
c_{21} & c_{22} & c_{23} \\
c_{31} & c_{32} & c_{33}
\end{array}
\right),
\end{align}
with
\begin{eqnarray}
c_{11}&=&-{10\over 3} a_{11}^2 a_{12}^2 + a_{12}^4 + a_{11}^4, \nonumber\\ 
c_{12}&=&4\sqrt{2\over 3} a_{11} a_{12}^3-4\sqrt{2\over 3} a_{11}^3 a_{12}, \nonumber\\ 
c_{13}&=&{8\sqrt{2}\over 3} a_{11}^2 a_{12}^2, \nonumber\\ 
c_{22}&=&a_{12}^4 + a_{11}^4-6a_{11}^2 a_{12}^2,  \nonumber\\ 
c_{23}&=&{4\over \sqrt{3}}a_{11}a_{12}^3-{4\over \sqrt{3}}a_{11}^3a_{12},\nonumber\\ 
c_{33}&=&-{2\over 3}a_{11}^2 a_{12}^2 + a_{12}^4 + a_{11}^4.
\end{eqnarray}
and $c_{21}=-c_{12}$, $c_{32}=-c_{23}$, and $c_{31}=c_{13}$.

For the quantum number of $K=6$ and $L=0$,
\begin{align}
R^{(ij)}_{4\times 4} = \left(
\begin{array}{cccc}
c_{11} & c_{12} & c_{13} & c_{14}\\
c_{21} & c_{22} & c_{23} & c_{24}\\
c_{31} & c_{32} & c_{33} & c_{34} \\
c_{41} & c_{42} & c_{43} & c_{44}
\end{array}
\right),
\end{align}
with
\begin{eqnarray}
c_{11}&=&a_{11}^6+7a_{11}^2a{12}^4-7a_{11}^4 a_{12}^2 - a_{12}^6\nonumber\\  
c_{12}&=&-2\sqrt{5}a_{11}a_{12}^5
-2\sqrt{5}a_{11}^5 a_{12}+{28\sqrt{5}\over 5}a_{11}^3 a_{12}^3,\nonumber\\  
c_{13}&=&-8 a_{11}^2 a_{12}^4 + 8a_{11}^4 a_{12}^2, \nonumber\\  
c_{14}&=&-{16\sqrt{5}\over 5}a_{11}^3 a_{12}^3, \nonumber\\ 
c_{22}&=&a_{11}^6+{67\over 5}a_{11}^2a_{12}^4-{67\over 5}a_{11}^4a_{12}^2-a_{12}^6, \nonumber\\ 
c_{23}&=&{32\over \sqrt{5}} a_{11}^3 a_{12}^3-{8\over \sqrt{5}}a_{11} a_{12}^5- {8\over \sqrt{5}}a_{11}^5 a_{12}, \nonumber\\ 
c_{24}&=&-{24\over 5}a_{11}^2 a_{12}^4+{24\over 5}a_{11}^4 a_{12}^2,\nonumber\\ 
c_{33}&=&a_{11}^6 - 7 a_{11}^4 a_{12}^2 + 7 a_{11}^2 a_{12}^4 - a_{12}^6,\nonumber\\ 
c_{34}&=&-{6\over \sqrt{5}} a_{11}^5 a_{12} + {4\over \sqrt{5}}  a_{11}^3 a_{12}^3 - {6\over \sqrt{5}}  a_{11} a_{12}^5,\nonumber\\ 
c_{44}&=&a_{11}^6 -{3\over 5} a_{11}^4 a_{12}^2 + {3\over 5} a_{11}^2 a_{12}^4 - a_{12}^6,
\end{eqnarray}
and $c_{21}=-c_{12}$, $c_{31}=c_{13}$, $c_{41}=-c_{14}$, $c_{32}=-c_{23}$, $c_{42}=c_{24}$, and $c_{43}=-c_{34}$.

\end{appendix}

\vspace{1cm}
\bibliographystyle{apsrev4-1.bst}
\bibliography{Ref}

\end{document}